# A combined NMR and DFT study of Narrow Gap Semiconductors: The case of PbTe

Robert E. Taylor[1], Fahri Alkan[2], Dimitrios Koumoulis[1], Michael P. Lake[1], Daniel King[1], Cecil Dybowski[2] and Louis-S Bouchard[1,3*]

[1]Dept. of Chemistry and Biochemistry, University of California, Los Angeles, CA  90095-1569  USA
[2]Dept. of Chemistry and Biochemistry, University of Delaware, Newark, DE  19761-2522  USA
[3]California NanoSystems Institute, UCLA, Los Angeles, CA 90095-1569 USA

*Corresponding author:  Louis-S Bouchard (email:  bouchard@chem.ucla.edu)

**Abstract**

In this study we present an alternative approach to separating contributions to the NMR shift originating from the Knight shift and chemical shielding by a combination of experimental solid-state NMR results and *ab initio* calculations. The chemical and Knight shifts are normally distinguished through detailed studies of the resonance frequency as function of temperature and carrier concentration, followed by extrapolation of the shift to zero carrier concentration.  This approach is time-consuming and requires studies of multiple samples.  Here, we analyzed $^{207}$Pb and $^{125}$Te NMR spin-lattice relaxation rates and NMR shifts for bulk and nanoscale PbTe.  The shifts are compared with calculations of the $^{207}$Pb and $^{125}$Te chemical shift resonances to determine the chemical shift at zero charge carrier concentration.  The results are in good agreement with literature values from carrier concentration-dependent studies.  The measurements are also compared to literature reports of the $^{207}$Pb and $^{125}$Te Knight shifts of *n*- and *p*-type PbTe semiconductors.  The literature data have been converted to the currently accepted shift scale.  We also provide possible evidence for the "self-cleaning effect" property of PbTe nanocrystals whereby defects are removed from the core of the particles, while preserving the crystal structure.

**Key Words**  [207]Pb NMR; [125]Te NMR; lead telluride; nanocrystals; self-cleaning effect; Knight shift; chemical shift; spin-lattice relaxation.

**Introduction**

Lead telluride is a narrow gap semiconductor long used for thermoelectric applications. Snyder and co-workers[1] recently provided an overview of the scientific and technological history of this material. The overview begins "during the Cold War and Space Race of the middle part of the 20th century" and concludes by discussing how band structure engineering to improve overall thermoelectric efficiency has recently been driving this field of research. Among the many ways to interrogate the state of semiconductors such as PbTe, nuclear magnetic resonance (NMR) of both nuclei has been extremely valuable in assessing the electronic state[2-7]. For example, the Knight shift[8], which probes the interaction of nuclear spins with conduction band carriers (electrons or holes), provided a direct measure of carrier concentration. This readout can be performed even on samples that are not amenable to transport studies. Such readouts could, in principle, aid in the development of novel materials which cannot be produced as high quality thin films or single crystals. This potentially important technological application is currently hampered, however, by our inability to separate Knight shift from chemical shift.

The Knight shift is typically measured relative to the resonance of the nucleus of interest in a "carrier-free sample"[3], thus the importance of the reference point. Nuclei such as [207]Pb experience strong chemical shielding contributions to the overall shift. These early studies report the Knight shifts as the magnetic field for the resonance of [207]Pb and [125]Te nuclei at a given radio frequency (RF). In these terms, the Knight shift, $K$, is given by[6]

$$K = (H_r - H_s)/H_s \qquad (1)$$

where $H_r$ is the magnetic field at which resonance of a sample free of charge carriers occurs, i.e. a sample whose resonance is governed only by the chemical shielding[9,10]. $H_s$ is the resonance field of a "real sample [...] at constant frequency"[3]. The chemical-shielding reference field, $H_r$, is usually determined indirectly by extrapolating the dependence of $H_s$ on carrier concentration (measured by some other technique) back to zero carrier concentration.

There are some challenges in determining the chemical shielding and expressing magnetic field values reported in the earlier literature in terms of the current convention for chemical shifts[11]

$$\delta \, (ppm) \;=\; \left\{ \frac{(sample - reference)}{reference} \right\} \times 10^6 \qquad (2)$$

One challenge is the accuracy with which the resonance field can be read from graphs in the older literature. Another issue arises from the temperature dependence of the chemical shifts and Knight shifts. The chemical shifts of diamagnetic lead compounds often have a strong dependence on temperature. As an example, the $^{207}$Pb resonance of lead nitrate is routinely used as a NMR thermometer because of the 0.76 ppm/K change of the resonance position with temperature[12-19]. Separating the temperature dependence of the Knight shift from that of the chemical shift can be done only for temperature studies with different samples of *n*- and *p*-type semiconductors. This procedure is time-consuming and strongly depends on the quality of the samples and their characterization by transport measurements. The reliability of transport measurements in turn depends on sample quality.

Previous experiments focused on $^{207}$Pb and $^{125}$Te NMR line positions as functions of carrier concentration and temperature; band structure calculations provided a qualitative understanding of the results and energy gap. The $^{207}$Pb and $^{125}$Te chemical shifts relative to the resonance of a standard material are dependent on the local structure and electronic environment.

As a result, chemical shift tensors provide qualitative trends that yield insights into the chemical bonding. The structure of PbTe is particularly simple, which allows one to compare experimentally-measured chemical shifts with first-principles calculations.

The nuclear magnetic relaxation of $^{207}$Pb and $^{125}$Te reflects the dynamics of fluctuating magnetic fields in the vicinity of the nucleus. In semiconductors, relaxation is frequently dominated by spin-flip scattering with the conduction charge carriers. As such, the relaxation behavior is correlated with the Knight shift. The relaxation behavior of both $^{207}$Pb and $^{125}$Te as a function of temperature confirms, as with the shifts, that the sample contains a distribution of electronic environments[20].

**Experimental and Theoretical Methods**

The NMR data were acquired with a Bruker DSX-300 spectrometer operating at a frequency of 62.79 MHz for $^{207}$Pb and 94.69 MHz for $^{125}$Te. A standard Bruker X-nucleus wideline probe with a 5-mm solenoid coil was used with a static polycrystalline sample of PbTe. The PbTe samples were obtained from Sigma Aldrich (avg. size: 84% smaller than 177 μm and 13% between 177 and 354 μm) and Alfa Aesar. The PbTe particle size was small enough to avoid RF skin-depth effects at these frequencies. The sample was confined to the length of the RF coil. The $^{207}$Pb $\pi/2$ pulse width was 4.5 μs, and the $^{125}$Te $\pi/2$ pulse width was 4 μs. We have verified that these RF pulses excited the entire NMR resonances. Magic-angle-spinning (MAS)[21-23] spectra were acquired on the same sample with a standard Bruker MAS probe using a 4-mm outside diameter zirconia rotor with a sample spinning rate of 10 kHz. The $^{207}$Pb and $^{125}$Te $\pi/2$ pulse widths for the MAS experiments were 4.4 μs. To minimize pulse ringdown effects, spectral data were acquired using a spin-echo sequence [$(\pi/2)_x - \tau - \pi)_y - \tau$ - *acquire*] with the echo delay, τ, set to 20 μs. Data for determining the spin-lattice relaxation times ($T_1$) were acquired with a saturation-recovery technique[24]. The $^{207}$Pb and $^{125}$Te chemical shift scales were calibrated using the unified Ξ scale[11], relating the nuclear resonance frequency to the $^1$H resonance of dilute tetramethylsilane in CDCl$_3$ at a frequency of 300.13 MHz. The reference for $^{207}$Pb is tetramethyllead (TML) and the reference for $^{125}$Te is dimethyltelluride. The two chemical shift scales were experimentally verified by acquiring spectra of lead nitrate[16-18] and telluric acid[25]. We also note the existence of alternative $^{125}$Te absolute chemical shielding scales and comparisons to density functional theory (DFT) predictions[26-28]. A relaxation delay of at least five times $T_1$ was used in each experiment to allow full recovery of the magnetization.

Quantum calculations were performed with the Amsterdam Density Functional theory (ADF) suite of software[29-31] on models of the local PbTe environment. Calculations employed the Becke88-Perdew86 generalized gradient approximation (BP86) functional[32-33] and a triple-zeta double polarized (TZ2P) basis set. The calculations included relativistic effects by the zero order regular approximation (ZORA) to the two-component Dirac equation with spin-orbit coupling[34-35]. NMR chemical shieldings were calculated with the NMR module[36] associated with the ADF program package. The $^{207}$Pb chemical shifts are reported relative to tetramethyllead (TML), whose isotropic chemical shift was calculated at the same level of theory. The $^{125}$Te chemical shifts are reported relative to dimethyltelluride.

**Results and Discussion**

**Chemical Shifts**

With modern NMR spectrometers, the $^{207}$Pb resonance frequency is typically measured in a magnetic field of a fixed strength. The measured NMR shift includes contributions from chemical shielding and the Knight shift. For diamagnetic insulators, the chemical shift of $^{207}$Pb can range up to 12,000 ppm relative to tetramethyllead[37]. Similarly, the range of chemical shifts of $^{125}$Te is approximately 5,000 ppm. The $^{207}$Pb Knight shifts are among the largest known shifts for nontransition materials, on the order of 1% for a carrier concentration of $10^{19}$ cm$^{-3}$ in *n*- and *p*-type PbTe semiconductors[5] (although the $^{125}$Te shift remains small in these materials). As mentioned previously, the contributions of the two components to the overall shift can be separated because the Knight shift is experimentally determined by measuring the resonance position relative to that of a "carrier-free sample"[3].

Lead telluride has been investigated with temperature-dependent NMR spectroscopy by several groups[2-7]. The theoretical challenge has been to explain the large Knight shift of $^{207}$Pb arising from both the *s*-like holes and *p*-like conduction electrons. As discussed by Sapoval and co-workers[4,5], the early theory suggested that only *s*-like electrons give rise to a large Knight shift. They noted that more comprehensive models with the full expression for hyperfine coupling indicated that large Knight shifts can also arise from the spin-orbit term for *p*-type electrons and that a dipolar field from the electron, "*even in cubic crystals*"[5], can arise due to the spin-orbit term.

These early studies typically reported Knight shifts as the magnetic field for the resonance of $^{207}$Pb nuclei at a given radiofrequency according to Eq. (1). These experiments were done in the swept-field/constant frequency mode. The resonance fields of $^{207}$Pb at a frequency of 7.5056 MHz as a function of the carrier concentration for a series of *n*- and *p*-type PbTe at 1.3 K have been measured and extrapolated back to a field of 8436 gauss for a hypothetical sample of zero carriers[2,5] to yield the chemical shift of PbTe at that temperature. A different study[3] reported the resonance fields of $^{207}$Pb at a frequency of 10.0 MHz as a function of the carrier concentration for a series of *n*- and *p*-type PbTe samples from 77 K to 465 K. This work indicated a field of 11,216 gauss as the zero-charge-carrier resonance field at 300 K. The combination of these two studies provides the temperature dependence of the chemical shift of PbTe.

As the resonance frequency is directly proportional to the magnetic field[8], the observed resonance frequency can be scaled to a different magnetic field. This process was used in many early studies for comparison of experimental results. To express these experimental results from earlier studies in terms of current conventions[11] and to compare them directly with the experimental results reported in the present work, we have converted these earlier results to the appropriate frequencies that would be observed in a magnetic field of 7.04925 T in which the protons of tetramethylsilane are observed at a frequency of 300.130 MHz. These frequencies at 7.04925 T were then converted to parts per million (ppm) relative to the reference material by use of Eq. (2) and the unified $\Xi$ scale as described in Ref. (11). (Examples of this conversion process are explicitly shown in the *Supporting Information* section.) Sapoval's extrapolation[4] for the $^{207}$Pb chemical shift of PbTe at 1.3 K can be expressed as -420.8 ppm. The variable-temperature study[3] yields a $^{207}$Pb chemical shift of 970.8 ppm at 300 K.

The $^{207}$Pb spectrum of a static polycrystalline sample of PbTe at 295 K is shown in Fig. 1(a). The resonance line shape is clearly asymmetric. Such an asymmetrical feature likely arises from interaction with conduction carriers[38-47]. In a prior study[48,49], asymmetric resonance lines were attributed to an inhomogeneous distribution of defects yielding different charge carrier concentrations in different microcrystalline regions, giving a range of Knight shifts for spins in these different environments. Specifically, the $^{207}$Pb Knight shift in the case of PbTe with low *p*-type carrier concentrations has been reported[5] to go as $n^{1/3}$, where *n* is the charge carrier concentration. The tensorial nature of the electron-nuclear hyperfine interaction, namely the contributions from dipolar and orbital terms in a material with *p*-band charge carriers, could also contribute to the asymmetry of the line shape[50].

In the $^{207}$Pb spectrum of the static sample shown in Fig. 1(a), the experimental resonance extends from roughly 1,600 ppm to about -500 ppm, with a maximum intensity at ca. 875 ppm. In Fig. 1(a), the spectrum of the sample obtained under magic-angle spinning (MAS) has essentially the same structure, but it is slightly shifted to higher frequencies, with the maximum intensity at ~1,000 ppm. If the least shielded areas of the sample correspond to lower carrier concentrations, the results of Fig. 1(a) suggest a chemical shift at 295 K of between 1,200 and 1,400 ppm. For comparison with the earlier literature (as converted for the present work), Sapoval and co-workers[2,5] report a chemical shift of -420.8 ppm at a temperature of 1.3 K. The later work by Senturia, *et al.*[3] gives a chemical shift of 970.8 ppm at 300 K. From these observations, we conclude that the $^{207}$Pb chemical shift of PbTe is in the range of 1,000 to 1,400 ppm, and may depend on how the experiment is carried out, including the analysis of the variation with charge carrier concentration[38,39,49].

These "experimental" $^{207}$Pb chemical shifts for PbTe, obtained from extrapolation to zero concentration, for the carrier density can be compared with the chemical shift calculated with DFT including relativistic effects through the zero order regular approximation (ZORA) that includes scalar and spin-orbit effects. For the lead chemical shifts, the model is a lead atom surrounded by an octahedral array of six tellurium atoms; six hydrogen atoms were included to compensate the major charge on the cluster. For chemical shift referencing, the lead chemical shielding of tetramethyllead (TML) was calculated at the same level of theory. The results in Table I indicate that the diamagnetic contribution to the chemical shielding of lead is virtually independent of the Pb-Te distance, reflecting the major contributions from lead core electrons. This relative constancy of the diamagnetic contribution has also been observed in calculations of the $^{207}$Pb chemical shifts of the lead dihalides and the lead tetrahalides[51]. The paramagnetic and spin-orbit contributions, on the other hand, depend on the Pb-Te bond distance, so that the total chemical shift is dependent on the Pb-Te bond distance, as can be seen by the results in the last column of Table I. The calculated shift of 1,550 ppm for a PbTe bond length of 323 pm, matching the bond length reported in the crystal structure[52,53], is in agreement with the shifts observed in the commercial PbTe sample containing native defects and impurities.

For the tellurium resonance, the line shape of the commercial sample is relatively symmetric, as can be seen in Fig. 1(b), with a mean resonance position of -1,176 ppm. The resonance covers a range from roughly -1,150 ppm to -1,225 ppm. In this case it is difficult to know the chemical shift by assuming it is at the "edge" of the pattern, but a rough estimate would be around -1,150 ppm. The earlier literature, in a study[3] of *n*- and *p*-type PbTe as a function of the carrier concentration, indicated a magnetic field of 9,689 gauss as the resonance field at 13.0 MHz for the $^{125}$Te chemical shift at 77 K. This corresponds to a chemical shift of -1,147 ppm at 77 K. The measured $^{125}$Te shift of -1,176 ppm suggests the sample is a *p*-type semiconductor, in agreement with the $^{207}$Pb result. According to Balz et al.[54], the shape of $^{125}$Te MAS spectra (e.g. Fig. 1(b)) could be explained using $J(^{207}Pb, ^{125}Te) = 2150$ Hz for various isotopomers of a TePb$_6$ species to account for the width and shoulders in the spectrum. The chemical shift of the $^{125}$Te resonance has been explained in Ref. (56).

A model to calculate the $^{125}$Te chemical shift is a cluster consisting of a tellurium atom surrounded by an octahedral array of six lead atoms. To terminate the structure, fluorine atoms are appended to the lead atoms. The chemical shift as a function of Pb-Te distance is given in Table II, reported as the shift relative to that of neat dimethyltelluride[11] calculated at the same level of theory. The predicted chemical shift changes with the internuclear distance, but not to the extent that the lead chemical shift changes with internuclear distance. Over the same range of bond distance, the lead shift changes roughly five times as rapidly as the tellurium chemical shift. This theoretical observation is consistent with the generally narrower $^{125}$Te band as compared to the changes of lead chemical shifts. The predicted $^{125}$Te chemical shift is -1,134 ppm for the measured bond length. This is in good agreement with the experimentally observed value of -1,176 ppm.

**Spin-lattice Relaxation**

The spin-lattice relaxation time, $T_1$, is a useful probe of electronic structure in semiconductors. The $^{207}$Pb and $^{125}$Te spin-lattice relaxation behaviors in saturation-recovery experiments have been measured as a function of temperature. A typical recovery of $^{207}$Pb nuclear longitudinal magnetization at 295 K is shown in Fig. 2, together with a stretched exponential function representing the "best fit" to the data. Each data point in the graph of Fig. 2 represents the integral of the NMR spectrum across the entire range of frequencies. Therefore, relaxation in regions with different Knight shifts are included, yielding a distribution of relaxation times. The data are not fit by a single exponential as would be expected if all nuclei in the sample recovered with the same rate constant. Figure 3 compares the fully relaxed $^{207}$Pb spectrum with a partially relaxed spectrum ($\tau$ = 10 ms) of the sample. It is clear from this figure that the relaxation is inhomogeneous, with the regions of the line shape at more negative shifts being relaxed faster than those at less negative shifts. Treating the most intense spectral feature around 875 ppm as a single isochromat yields a $^{207}$Pb $T_1$ of approximately 27 ms, while the more rapidly recovering part in the range 700 ppm to -500 ppm has a $T_1$ around 12.7 ms. As the rate of relaxation, $1/T_1$, is known to be proportional to the carrier density[20,38,48], the variation of $T_1$ across the line shape suggests that the line shape arises from an inhomogeneous distribution of defects, creating domains with differing carrier concentrations[46,47]. The shorter $T_1$ and the increased Knight shift to lower frequency are consistent with an increased carrier concentration[20,46]. This shift to lower frequency of this sample appears to place it in the range of resonances for *p*-type PbTe semiconductors[3], as mentioned above.

As a result of the distribution of defects, the $^{207}$Pb spin-lattice relaxation data is best characterized by a Kohlrausch (stretched-exponential) function

$$M(t) = M_0(1 - \exp\left(-\frac{t}{T_1}\right)^\beta). \quad (3)$$

The application of this empirical model to physical systems, including NMR relaxation, is well established[56]. It is a natural choice for PbTe, which is characterized by an inhomogeneous distribution of carrier concentrations and defects. The results of this analysis of the $^{207}$Pb relaxation rate are shown in Fig. 4. The fitting of this plot gives an activation energy of 166 meV (16.02 kJ/mol). This value is virtually identical to the 160 meV (15.4 kJ/mole) activation energy for processes that induce relaxation of $^{205}$Tl in semiconducting Tl$_2$Se [57].

The $^{125}$Te spin-lattice relaxation ($T_1$) data from a saturation-recovery experiment at 295 K are shown in Fig. 2, as presented next to $^{207}$Pb results. As is the case for $^{207}$Pb, the relaxation clearly cannot be described as an exponential approach to equilibrium. The result of fitting these data to a stretched exponential function is shown in Fig. 5. (For the interested reader, a biexponential analysis of the same data is provided in the *Supporting Information* section.) In contrast to the results obtained by fitting a stretched-exponential to the $^{207}$Pb relaxation data, the $^{125}$Te results do not follow Arrhenius behavior with temperature. The low-temperature behavior indicates that at temperatures below about 250 K, another relaxation mechanism becomes sufficiently effective that it dominates relaxation. The limited data in this region indicate an activation energy of about 62.72 meV for this process. At higher temperatures a second relaxation mechanism becomes dominant, characterized by an activation energy of 159.24 meV. This second mechanism has, within experimental error, the same activation energy as was observed through the $^{207}$Pb spin-lattice relaxation. It is clear that relaxation of $^{125}$Te in this temperature region and $^{207}$Pb are effected by the same processes, an interaction with the conduction charge carriers, but that $^{125}$Te experiences another mechanism at low temperatures that is not seen to affect the $^{207}$Pb relaxation.

**Temperature-Dependent Changes in Line Shapes**

As discussed in the Introduction, there are challenges in accurately determining the "experimental" chemical shift from these earlier studies of PbTe. The challenges include the extrapolation to or estimation of the resonant magnetic field for a sample of zero charge carrier concentration, reading such values from graphs in the literature, and dealing with the effects of temperature upon the shifts. There are also challenges in accurately calculating the chemical shifts of such heavy nuclei that result in significant uncertainty in the predicted chemical shifts. As a result, the assignment of the PbTe used in this study as a *p*-type semiconductor based on comparisons with the $^{207}$Pb and $^{125}$Te "experimental" chemical shifts obtained from the shifts of samples as a function of charge carrier concentration, while strongly suggestive, is still somewhat tenuous. The calculated chemical shifts, also strongly suggestive, do not clearly resolve the issue. The $^{207}$Pb shift of this sample as a function of temperature, shown in Fig. 6, exhibits the same behavior as that of the *p*-type samples given in Fig. 3 of Ref. (3). Increasing the temperature of *p*-type samples of PbTe moves the resonant magnetic field to lower values (corresponding to increases in frequencies at a fixed magnetic field). The opposite behavior is observed for *n*-type samples. This temperature dependence of the $^{207}$Pb shift confirms the sample used in this study is indeed a *p*-type semiconductor.

Although the $^{125}$Te chemical shifts at 77 K have been reported as a function of carrier concentration[3], these early studies of PbTe did not provide any results for the $^{125}$Te chemical shift as a function of temperature for comparison with those obtained in this study. The dependence of the $^{125}$Te shift on temperature for the *p*-type sample of PbTe used in this work is shown in Fig. 7. Periodic checks of the sample integrity were made during the variable temperature (VT) studies. This was done by acquiring a one-dimensional $^{207}$Pb spectrum of the sample at 295 K after various stages of the VT study. These results are shown in Fig. 8. A change in the spectrum appeared after elevating the temperature to 423 K with the $^{207}$Pb spectral features moving to a slightly higher NMR frequency. This change is likely due to some annealing of the native defects. The effect is to move the resonance position closer to that of the true chemical shift, as shown in Fig. 8. The annealing temperature used in our study is lower than those used in the annealing study of Hewes *et al.*[58]. Due to the high sensitivity of the NMR experiment to sample defects, homogeneity and carrier concentration (see also, Ref. 57), accuracy in such estimates may require the use of one or more annealing cycles. The lack of sample annealing may lead to systematic errors in the estimation of the chemical shift comparable to that shown in Fig. 10. In addition to the observed $^{207}$Pb shift, the changes upon annealing will also manifest themselves in the $^{207}$Pb spin-lattice relaxation time.

Another source of systematic errors arises from the powdering of the samples and can alter the NMR shift (and consequently, the spin-lattice relaxation rate). The early NMR studies[2-

[7] of semiconductors usually determined the carrier concentrations on single crystal specimens and then powdered the samples for the NMR measurements. Hewes et al.[6] noted that the powdering process can alter the carrier concentration. For lead telluride, differences in carrier concentration were noted between those samples cleaved with a razor blade and those crushed with a mortar and pestle[6,58]. The annealing study[58] of *n*- and *p*-type PbTe should generally help in determining error propagation arising from changes in carrier concentration, as carrier concentration as function of annealing temperature in the range of 452 K and 1,000 K are reported. In previous studies on PbTe samples ball milled for 60 min[59,60], the crystallinity was preserved, but carrier concentration decreased. The decrease in carrier concentration in nanocrystalline samples is referred as the "self-cleaning" process of the bulk[61-64] whereby defects are pushed toward the surface. The nanoparticles possess fewer defects than the corresponding bulk material. Because of self-cleaning in nanoparticles, ball-milling would be expected to decrease spin lattice relaxation rate ($1/T_1$), as shown in Fig. 5 for the finest ball milled PbTe (83 nm). The self-cleaning process is directly related to the observed increase in the activation energy in the high and low temperature regime (<250 K). In Fig. 9, powder x-ray diffraction (PXRD) spectra for powdered PbTe ingots (Alfa Aesar, Ward Hill, MA) mortar and pestle versus ball milled (83 nm, 60 min. ball milling) samples confirm that crystallinity is indeed preserved. However, the carrier concentration in these samples appears to change, as evidenced by the increased $T_1$ time with decreasing particle size (Table III).

Atomic self-diffusion of Pb and Te has been investigated[65] in lead telluride at temperatures in the range 793 - 1,000 K with use of radioactive tracers. The self-diffusion coefficients reported were quite high: $2.9 \times 10^{-5}$ cm$^2$/sec for Pb and $2.7 \times 10^{-6}$ cm$^2$/s for Te. While the activation energy for diffusion of Pb was dependent upon the stoichiometry of the sample, the lowest measured activation for both Pb and Te at 100 kJ/mole is still much higher than the activation energies found in the current investigation. Thus, the relaxation mechanism in the high-temperature regime cannot be due to atomic diffusion. The likely mechanism involves interactions of nuclear moments with thermally activated charge carriers[46,47,57]. The low-temperature process that affects the $^{125}$Te relaxation can be, like for TlTaS$_3$, due to spin diffusion to localized electron spins.

Magic-angle spinning (MAS) does not lead to a significant narrowing of the $^{207}$Pb resonance, as shown in Fig. 1(a). The shift to higher frequency is consistent with an increased temperature in the MAS rotor resulting from the sample spinning at 10 kHz[38,41,66]. A measurement of the $^{207}$Pb NMR shift of lead nitrate[12-19] at an ambient temperature of 295 K indicates a sample temperature of 306 K at this rotation rate. However, the magnitude of the change in the shift at 140 ppm for the MAS spectrum is much larger than that expected from frictional heating of the MAS rotor alone. Such an effect of similarly large shifts in MAS spectra of conductive samples has been explained[66] for spectra of CuI. This shift to higher frequency for the $^{207}$Pb resonance of a static sample of lead telluride as the temperature is increased is demonstrated in Fig. 6. MAS increases the magnitude of this shift in conductive samples.

MAS also does not substantially narrow the $^{125}$Te resonance, as shown in Fig. 1(b). The rotation rate of 10 kHz is larger than the $^{125}$Te static line width of 4.7 kHz. As MAS does not result in a manifold of spinning sidebands in either the $^{207}$Pb or $^{125}$Te MAS spectra, the broadening of the spectral resonance is not due solely to an orientational dependence of chemical shifts (or Knight shifts). While Figure 7 shows that the $^{125}$Te resonance of the static sample exhibits the same shift to increasing frequency as the temperature is raised as observed with the $^{207}$Pb spectra shown in Figure 6, the $^{125}$Te MAS spectrum in Figure 1(b) shows a slight narrowing of the resonance with a shift of the first moment of the resonance to lower frequency. This likely arises due to the averaging of anisotropic interactions.

**Conclusions**

Unless samples of zero Knight shift (i.e. with low carrier concentration) are available as reference, the chemical and Knight shifts are normally distinguished through detailed studies of the resonance frequency as function of temperature and carrier concentration, followed by extrapolation of the shift to zero carrier concentration. In order to overcome this time-consuming approach we presented here a combined NMR and DFT method to separate these two contributions. A second advantage of the approach is it enables studies of specific samples with fixed physical properties whose carrier concentrations cannot easily be altered. Calculations of the chemical shift of PbTe based on model structures are in agreement with experiment. Experiments confirm that PbTe consists of a distribution of crystallites having differing densities of charge carriers, resulting in an asymmetric line shape due to a distribution of Knight shifts in various domains of the sample. Our results suggest that the sample studied was *p*-type, in agreement with previous carrier concentration-dependent studies of *n*- and *p*-type PbTe semiconductor.

Relaxation studies show that the sample exhibits dynamics that arise from a range of charge carrier densities. In the high-temperature regime, relaxation is governed by interaction with thermally activated charge carriers. For $^{125}$Te at temperatures below about 250 K, a second mechanism dominates spin-lattice relaxation, which we postulate may be due to spin diffusion to isolated electron spins. The low temperature regime is also suggestive that a second mechanism affects $^{207}$Pb relaxation. Thus, interaction with both mobile carriers and fixed centers contributes to relaxation in PbTe.

The literature data on PbTe have been converted to the currently accepted shift scale. We also provide possible evidence for the "self-cleaning effect" property of PbTe nanocrystals whereby defects are removed from the core of the particles, while preserving the crystal structure.

**Acknowledgments**

This material is based upon work supported by the National Science Foundation through Grant CHEM0956006 to C.D. The work at UCLA (L.-S. B.) was supported by Defense Advanced Research Project Agency (DARPA), award no: N66001-12-1-4034. L.-S. B. acknowledges useful discussions with Mercouri G. Kanatzidis.

**Tables**

**Table I. Calculated Lead Chemical Shieldings for $[PbTe_6H_6]^{-4}$**

| Pb-Te distance (pm) | $\sigma_{dia}$ (ppm) | $\sigma_{para}$ (ppm) | $\sigma_{SO}$ (ppm) | $\sigma_{iso}$ (ppm) | $\delta_{iso}$ (ppm) |
|---|---|---|---|---|---|
| **298** | 9950 | -4960 | 2867 | 7857 | -513 |
| **303** | 9950 | -5182 | 2830 | 7599 | -256 |
| **308** | 9950 | -5463 | 2785 | 7273 | 71 |
| **313** | 9950 | -5815 | 2734 | 6870 | 474 |
| **318** | 9950 | -6249 | 2683 | 6384 | 959 |
| **323** | 9950 | -6794 | 2636 | 5793 | 1550 |
| **328** | 9950 | -7442 | 2606 | 5114 | 2229 |

**Table II. Calculated Tellurium Chemical Shieldings for $[TePb_6F_{18}]^{-8}$**

| Pb-Te distance (pm) | $\sigma_{dia}$ (ppm) | $\sigma_{para}$ (ppm) | $\sigma_{SO}$ (ppm) | $\sigma_{iso}$ (ppm) | $\delta_{iso}$ (ppm) |
|---|---|---|---|---|---|
| **308** | 5311 | -2499 | 1357 | 4169 | -1070 |
| **313** | 5311 | -2559 | 1439 | 4191 | -1093 |
| **318** | 5311 | -2628 | 1529 | 4212 | -1113 |
| **323** | 5311 | -2705 | 1626 | 4232 | -1134 |
| **328** | 5311 | -2792 | 1733 | 4252 | -1153 |
| **333** | 5311 | -2891 | 1846 | 4266 | -1167 |

**Table III. Comparison of $T_1$ relaxation times for lead telluride ball milled for various times.** The average particle size of nanocrystalline powders was estimated from PXRD data using the Scherrer formula.

| vendor | powdering method | milling time | average nanocrystal size | $T_1$ | Kohlrausch exponent (β) |
|---|---|---|---|---|---|
| Sigma-Aldrich | used as-is (mesh140) | n/a | n/a | 2.6 s | 0.80 |
| Alfa Aesar | mortar & pestle | n/a | n/a | 6.8 s | 0.44 |
| Alfa Aesar | ball milled | 1 min. | 326 nm | 6.7 s | 0.53 |
| Alfa Aesar | ball milled | 5 min. | 151 nm | 12.7 s | 0.55 |
| Alfa Aesar | ball milled | 15 min. | 131 nm | 12.7 s | 0.55 |
| Alfa Aesar | ball milled | 30 min. | 98 nm | 12.7 s | 0.55 |
| Alfa Aesar | ball milled | 60 min. | 83 nm | 12.7 s | 0.55 |

**Figures and Captions**

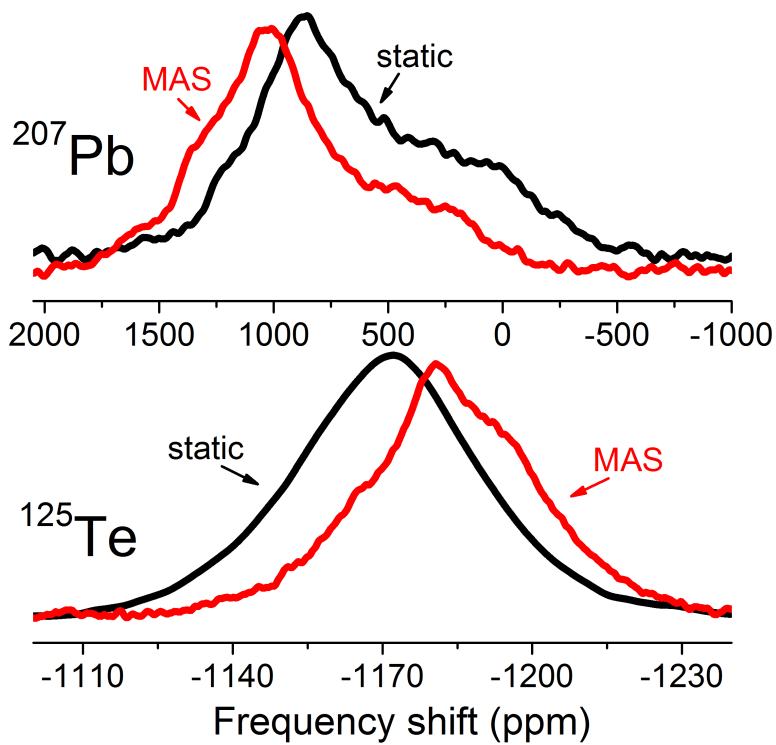

**Figure 1.** (a) $^{207}$Pb spectrum (black online, A) of static polycrystalline PbTe at 295 K. The $^{207}$Pb MAS spectrum is shown below (red online, B). (b) $^{125}$Te spectrum (black online, A) of static polycrystalline PbTe at 295 K. The $^{125}$Te MAS spectrum is shown below (red online, B).

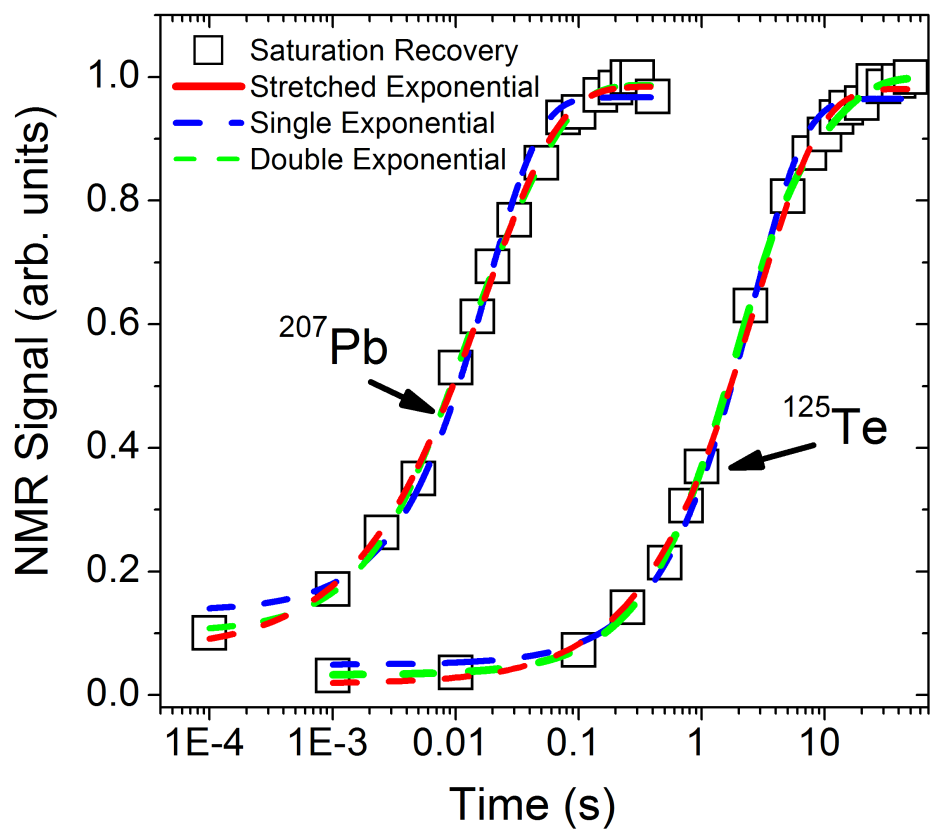

**Figure 2.** $^{207}$Pb and $^{125}$Te saturation recovery of PbTe. The data represent integral intensities of the entire resonance. The dashed line is a fit to a single-exponential (blue) and to a double exponential saturation-recovery model (green). The red line is a fit to a stretched-exponential saturation-recovery model.

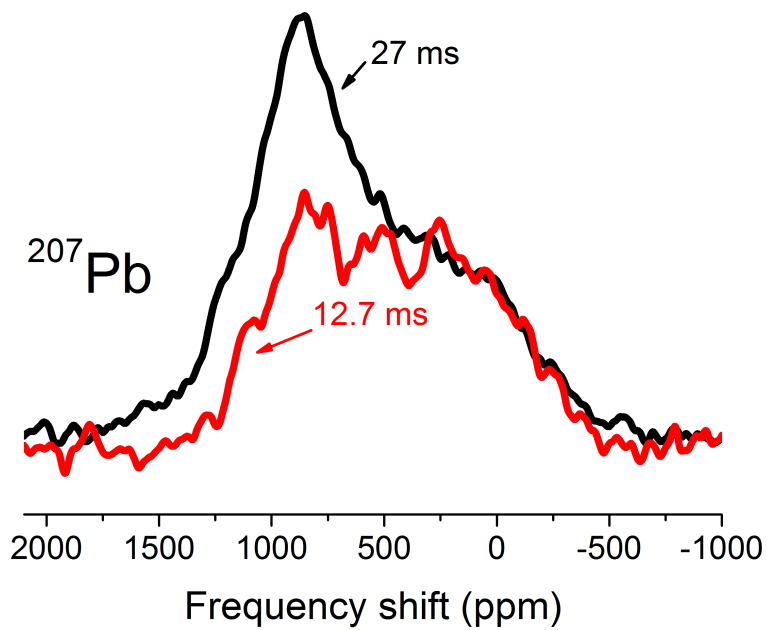

**Figure 3.** Comparison of the [207]Pb spectrum obtained in the saturation-recovery experiment with a delay after saturation of 10 ms (red on line) and for a fully relaxed spectrum (black on line) of PbTe at 295 K. Relaxation recovery is not uniform across the resonance. The regions at more negative shifts relax more quickly (12.7 ms) than do those at more positive shifts (27 ms, fully relaxed).

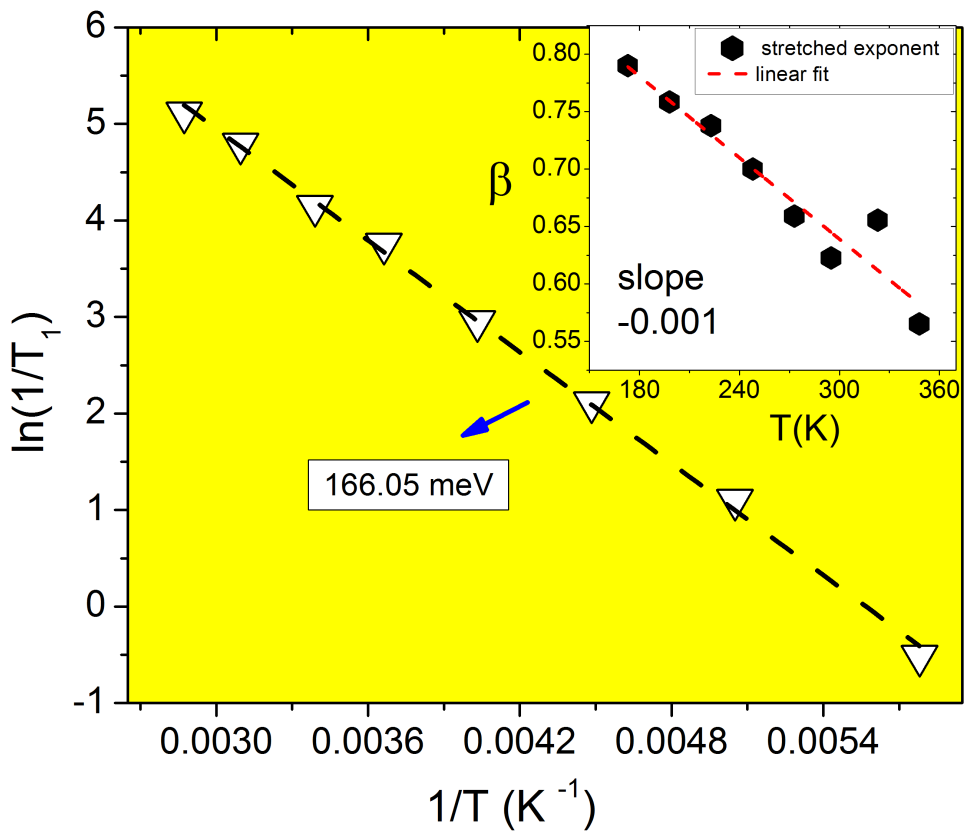

**Figure 4.** (a) The natural logarithm of the average $^{207}$Pb spin-lattice relaxation rate, as a function of the inverse temperature; (b) the exponent (β) obtained from a stretched-exponential fit plotted as a linear function of the temperature.

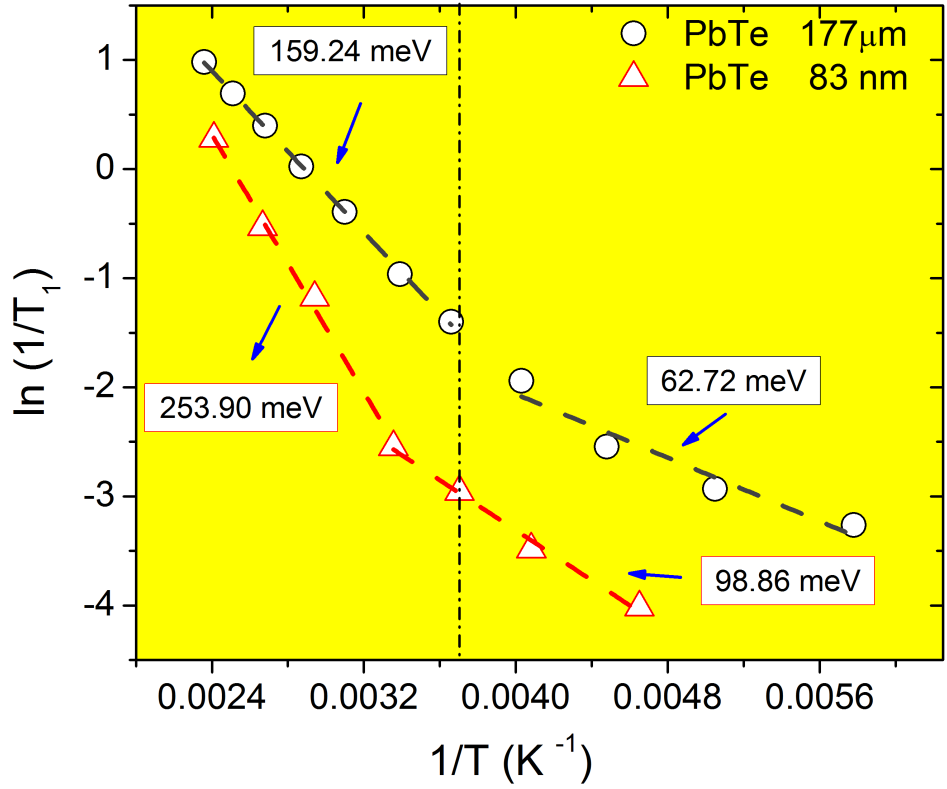

**Figure 5.** The natural logarithm of the [125]Te spin-lattice relaxation rate for the case of PbTe (177 μm, black open circles) and PbTe (83 nm, red open triangles), as a function of the inverse temperature. Below 250 K, a second mechanism dominates the spin lattice relaxation in both cases.

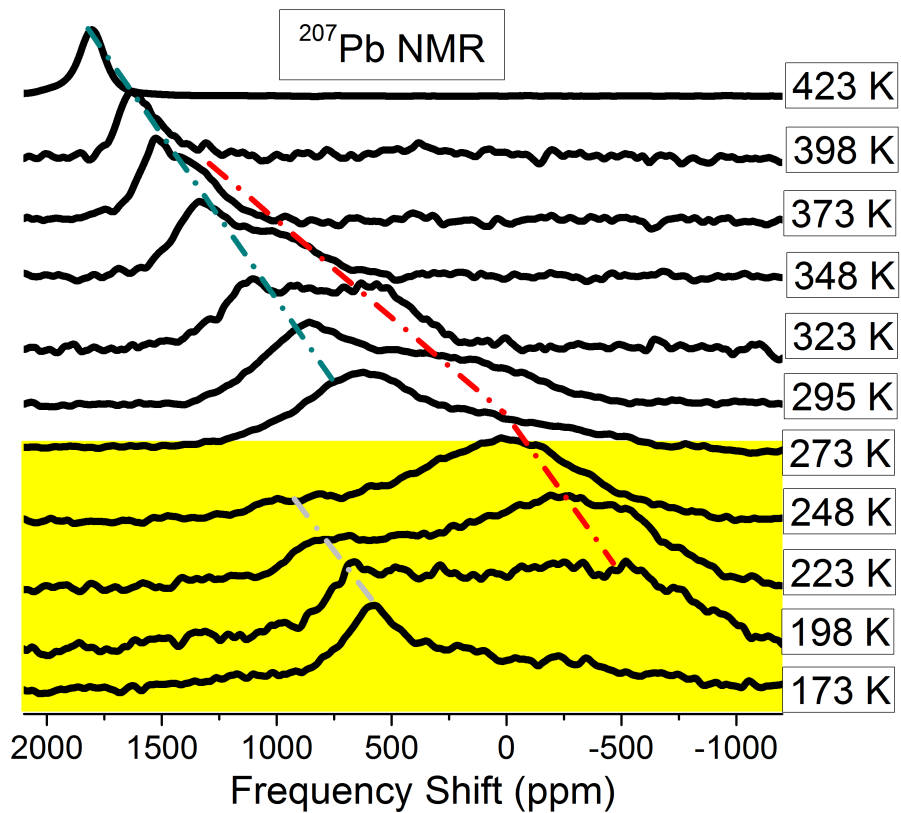

**Figure 6.** $^{207}$Pb NMR spectra of static polycrystalline PbTe from 173 K (bottom) to 423 K (top). The spectrum shape is changing as the temperature increases. Regions with different carrier concentrations shift differently with temperature (as shown by dashed lines). Above 250 K (white regime) the spectrum shifts to higher frequencies accompanied by a reduction in the spread of the spectrum. These two characteristics suggest an annealing of the native defects as the temperature increases.

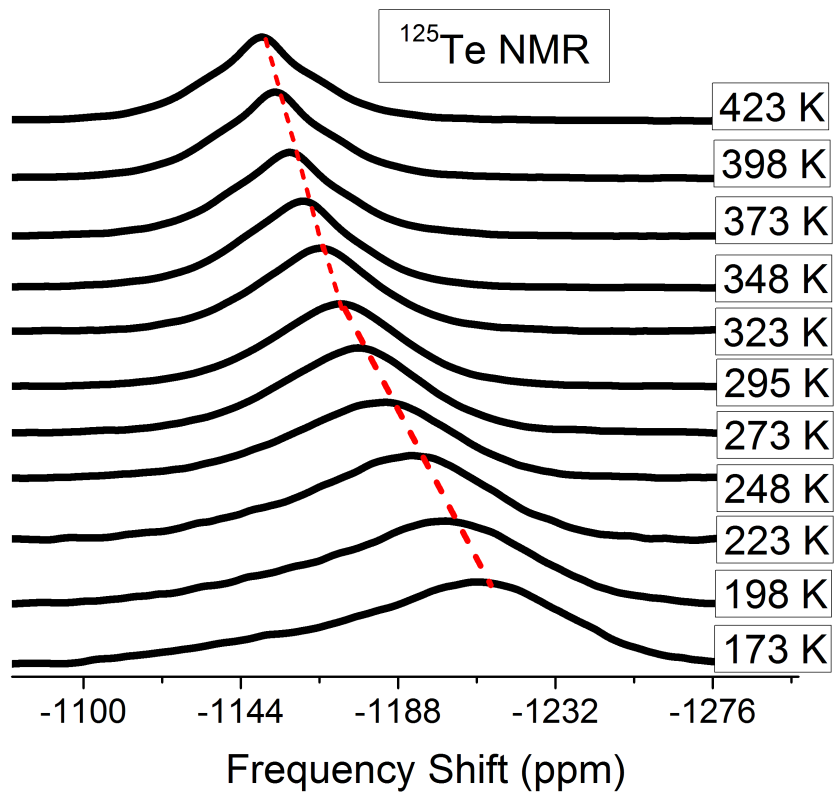

**Figure 7.** $^{125}$Te spectra of static polycrystalline PbTe from 173 K (bottom) to 423 K (top). With increasing temperature the observed positive shift of spectra is accompanied by a narrowing of the linewidth and by an alteration of $^{125}$Te spin-lattice relaxation mechanism.

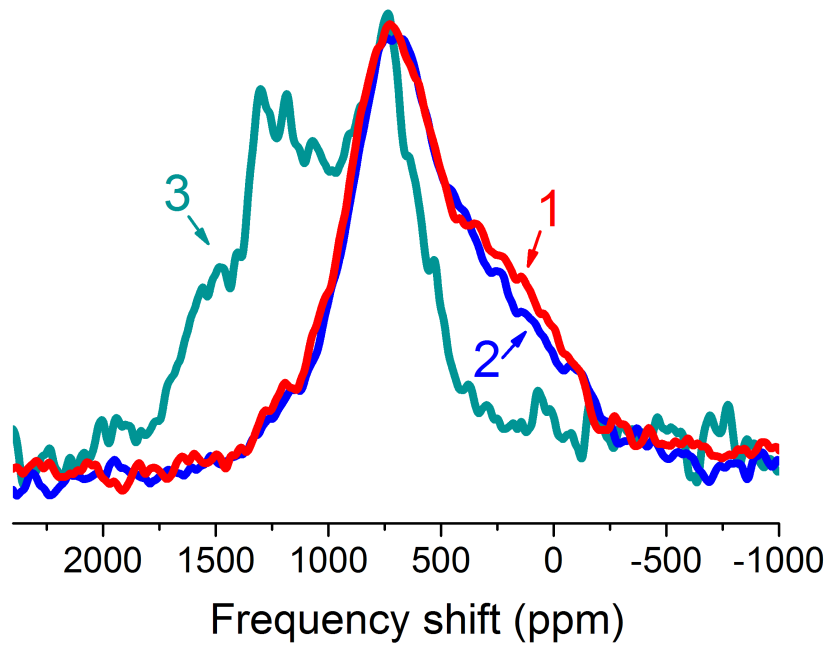

**Figure 8.** $^{207}$Pb spectra of static polycrystalline PbTe acquired at ambient temperature. The spectrum 1 (red on line, 1) is prior to any variable temperature experiments. The next spectrum (blue on line, 2) is at 295 K after acquiring data at 273 K and then up to and including 398 K. The spectrum 3 (dark cyan on line, 3) was acquired at 295 K after first heating the sample to 423 K. No further changes in the ambient temperature $^{207}$Pb spectrum were observed after the saturation-recovery experiments at any of the other lower temperatures.

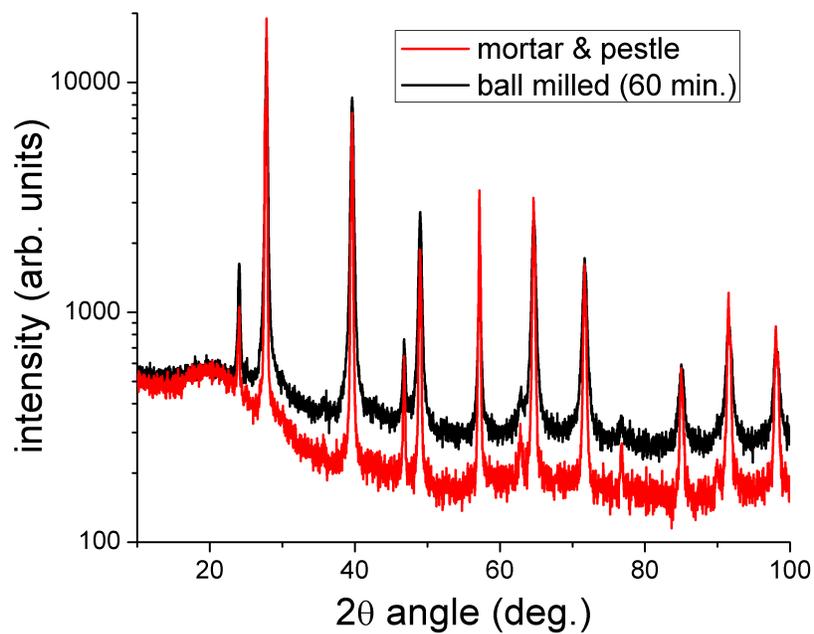

**Figure 9.** PXRD spectra for PbTe powders from mortar and pestle versus ball milled processes confirm that the samples preserve their crystallinity. For the ball-milled (60 min.) sample, the average particle size was estimated from the PXRD data to be 83 nm, using the Scherer formula.